\documentclass[12pt]{iopart}

\usepackage{iopams}  
\usepackage{cite}
\usepackage{cleveref}
\usepackage{hyperref}
\usepackage{graphicx}

\usepackage[dvipsnames]{xcolor}
\renewcommand{\theequation}{\thesection.\arabic{equation}}
\renewcommand{\theenumi}{(\thesection.\alph{enumi})}

\begin{document}

\title[On (super)integrable systems in magnetic fields with non-subgroup type integrals]{On rotationally invariant integrable and superintegrable classical systems in magnetic fields with non-subgroup type integrals}

\author{S Bertrand and L \v{S}nobl}
\address{Czech Technical University in Prague, Faculty of Nuclear Sciences and Physical Engineering, Department of Physics, B\v{r}ehov\'a 7, 115 19 Prague 1, Czech Republic}
\ead{bertrseb@fjfi.cvut.cz, Libor.Snobl@fjfi.cvut.cz}

\begin{abstract}
The aim of the present article is to construct quadratically integrable three dimensional systems in non-vanishing magnetic fields which possess so-called non-subgroup type integrals. The presence of such integrals means that the system possesses a pair of integrals of motion in involution which are (at most) quadratic in momenta and whose leading order terms, that are necessarily elements of the enveloping algebra of the Euclidean algebra, are not quadratic Casimir operators of a chain of its subalgebras. By imposing in addition that one of the integrals has the leading order term $L_z^2$ we can consider three such commuting pairs: circular parabolic, oblate spheroidal and prolate spheroidal. We find all possible  integrable systems possessing such structure of commuting integrals and describe their Hamiltonians and their integrals.

We show that our assumptions imply the existence of a first order integral $L_z $, i.e. rotational invariance, of all such systems. As a consequence, the Hamilton--Jacobi equation of each of these systems with magnetic field separates in the corresponding coordinate system, as it is known to be the case for all quadratically integrable systems without magnetic field, and in contrast with the subgroup type, i.e. Cartesian, spherical and cylindrical, cases, with magnetic fields.

We also look for superintegrable systems within the circular parabolic integrable class. Assuming the additional integral to be first order we demonstrate that only previously known systems exist. However, for a particular second order ansatz for the sought integral ($L^2+\ldots$) we find a minimally quadratically superintegrable system. It is not quadratically maximally superintegrable but appears to possess bounded closed trajectories, hinting at hypothetical higher order superintegrability.
\end{abstract}

\pacs{02.30.Ik,45.20.Jj}
\ams{37J35,78A25}

\vspace{2pc}
\noindent{\it Keywords}: integrability, superintegrability, classical mechanics, magnetic field.

%\submitto{\jpa}

%\today

\maketitle

\section{Introduction}\label{SecIntro}\setcounter{equation}{0}
In this paper we consider integrability and superintegrability of three dimensional (3D) mechanical systems in a magnetic field. We focus on systems related to three types of coordinates: circular parabolic, oblate spheroidal and prolate spheroidal.

The study of superintegrable systems, i.e. integrable systems possessing additional integrals of motion, in three spatial dimensions dates back to the pioneering work of Makarov, Smorodinsky, Valiev and Winternitz \cite{MSVW}. They found all systems without magnetic field that allow a pair of commuting integrals of motion which are (at most) second order in the momenta, established that they exactly coincide with 11 classes of systems known to be separable in some orthogonal coordinate system \cite{Eisenhart34,Eisenhart48} and studied some superintegrable subclasses. The research on second order superintegrability in 3D was completed by Evans \cite{Evans90}, and Kalnins and Miller et al \cite{KWMP,KKM07}. Higher order integrals were also subsequently considered, see e.g. \cite{KKM05,KKM07o,VE08,TD11} and others.

A natural extension of these results is to consider a similar problem, i.e. quadratic integrability and superintegrability, in the presence of the magnetic field. These questions were first adressed in two dimensions, starting with the paper \cite{DGRW} and followed up by many others, e.g. \cite{BW04,MW00,PR05,Pucacco04}. The question of superintegrability for the 3D problem was first considered in the particular case of magnetic monopole in \cite{MC10,LMV91}. A systematic approach to the 3D problem was initiated in \cite{MSW15}, followed up by a series of papers \cite{MS17,MSW18,MS18}. Related problems of superintegrability in relativistic mechanics were  also recently considered, with \cite{Ilderton18,IS18,HI17} and without \cite{AHI18} magnetic fields.

We should also recall that a related (but in the case of non-vanishing magnetic field in general non-equivalent) problem is the construction of systems whose Hamilton--Jacobi equation separates in some orthogonal coordinate system. This question was addressed by Shapovalov and Meshkov in the quantum case \cite{SBM72} and by Benenti, Chanu and Rastelli in the classical case\cite{BCR01}.

In order to study superintegrability one needs to first know the integrable systems. As was found first by Zhalij in \cite{Zhalij15} for Cartesian type integrals, the solution to this problem is significantly more demanding in the presence of magnetic fields and may lead to significantly increased number of non-equivalent cases. That study of integrable systems with Cartesian type integrals, i.e. integrals which would correspond to separation in Cartesian coordinates in the absence of the magnetic field, was followed by the spherical type \cite{MSW18} and also cylindrical type (work in progress). All these classes share one important property: the leading order terms of their commuting quadratic integrals of motion are second-order Casimir operators of subgroups in some subgroup chain
\begin{equation}
G \supset \tilde{G} \supset G_M
\end{equation}
where $G$ is the Euclidean group and $G_M$ is its maximal Abelian subgroup. The corresponding orthogonal coordinate systems (i.e. Cartesian, spherical, cylindrical) are called subgroup type coordinates \cite{KMW76,MPW81}. It was observed even in two spatial dimensions without magnetic field that systems separating in subgroup type and non-subgroup type coordinates have different properties, e.g. the existence of the so-called exotic potentials is restricted to subgroup-type systems \cite{MW08,Gravel04,Marquette10,MSW17,ELW17}. 

\medbreak

Thus we consider it appropriate to investigate the question of integrability and superintegrability for at least some classes of non-subgroup type integrals. A natural selection of classes from the point of view of potential physical interest is to study the non-subgroup type coordinates such that one of the integrals has leading order term of the form $L_z^2$, i.e. some kind of rotational invariance is present in the system. There are three classes of such coordinate systems, each in turn corresponding to one structure of the leading order terms in the integrals:
\begin{itemize}
\item circular parabolic, with the integrals of the form $L_xp_y-L_yp_x+\ldots$ and $L_z^2+\ldots$,
\item prolate spheroidal, with the integrals of the form $L^2+a^2(p_x^2+p_y^2)+\ldots$ and $L_z^2+\ldots$, and  
\item oblate spheroidal, with the integrals of the form $L^2-a^2(p_x^2+p_y^2)+\ldots$ and $L_z^2+\ldots$
\end{itemize}
These are the three classes of systems which we shall consider in the following and compare their properties with the ones of the systems possessing subgroup-type quadratic integrals.

\medskip

The paper is organized as follows. In section \ref{SecGen}, we present the notation used throughout the article and recall some basic facts about (super)integrability and magnetic fields. In section \ref{SecInt}, we investigate circular parabolic-type integrability for classical 3D Hamiltonian systems admitting non-zero magnetic fields. The determining equations are provided and solved for such a system in the circular parabolic coordinates under the assumption that the leading order terms of the integrals of motion correspond to separation of the Hamilton--Jacobi equations in the circular parabolic coordinate system when the magnetic field is absent. In section \ref{SecOblate}, we present as a result the non-zero magnetic field, the Hamiltonian system and its integrals of motion for which the leading order terms correspond to separation of variables of the Hamilton--Jacobi equations in the oblate spheroidal coordinates. Similarly, in section \ref{SecProlate}, we present analogue results but for the prolate spheroidal coordinates. In section \ref{SecSup}, we look in a systematic way for all additional first order integrals for the circular parabolic case in order to get superintegrable cases. In section \ref{SecL2}, we investigate the possibility of an additional integral of motion with a leading order term $(L^A)^2$, where a new component in the magnetic field appears. Some conclusions and future perspectives are discussed in section \ref{SecConc}.

\section{Essentials and notation about 3D (super)integrability with magnetic fields}\label{SecGen}
We are considering classical 3D Hamiltonian systems that admit a static electromagnetic field, i.e.
\begin{equation}
H=\frac{1}{2}\left(\vec{p}+\vec{A}(\vec{x})\right)^2+W(\vec{x}),\label{genH}
\end{equation}
where $\vec{p}$ is the momentum, $\vec{A}(\vec{x})$ is the vector potential depending on the position $\vec{x}$ and $W(\vec{x})$ is the scalar potential. The mass is set to 1 and the electric charge is assumed to be $-1$ in our units (implicitly expecting the considered particle to be an electron). The vector potential $\vec{A}(\vec{x})$ can be seen as a 1-form, e.g.
\begin{equation}
A=A_x(x,y,z)dx+A_y(x,y,z)dy+A_z(x,y,z)dz,\label{Agen}
\end{equation}
in the Cartesian coordinates. The magnetic field $\vec{B}(\vec{x})$ can be computed from the vector potential (\ref{Agen}) by taking its exterior derivative, i.e.
\begin{eqnarray}
\hspace{-2.5cm}B=\, dA&\hspace{-1cm}=\left(\partial_yA_z-\partial_zA_y\right)dy\wedge dz+\left(\partial_zA_x-\partial_xA_z\right)dz\wedge dx+\left(\partial_xA_y-\partial_yA_x\right)dx\wedge dy\nonumber\\
&\hspace{-1cm}=B_xdy\wedge dz+B_ydz\wedge dx+B_zdx\wedge dy.
\end{eqnarray}
The vector potential is defined up to a gauge transformation $\widetilde{A}=A+\nabla F$. Since we shall be interested in a static situation, i.e. $\vec{A}$ and $W$ are time independent, we also assume that the gauge transformation is time independent, and consequently $\widetilde{W}=W$.

We will be focusing on integrable Hamiltonian systems for which at least one of the components of the magnetic field $B_x$, $B_y$ or $B_z$ is not zero. For a 3D classical Hamiltonian system to be integrable in the sense of Liouville, it must possess two integrals of motion $X_1$ and $X_2$, which Poisson-bracket commute with the Hamiltonian $H$, i.e.
\begin{equation}
\lbrace X_i,H\rbrace=0,\qquad\mbox{where}\qquad \lbrace f,g\rbrace=\sum_{j=1}^{3}\frac{\partial f}{\partial x_j}\frac{\partial g}{\partial p_j}-\frac{\partial g}{\partial x_j}\frac{\partial f}{\partial p_j}.
\end{equation}
We also require that the integrals of motion $X_1$ and $X_2$ are in involution, that is $\lbrace X_1,X_2\rbrace=0$. In addition, the integrals of motion $X_i$ together with the Hamiltonian $H$ must be functionally independent, i.e. the matrix
\begin{equation}
\left[\frac{\partial(H,X_1,X_2)}{\partial(x_j,p_k)}\right]
\end{equation}
is of maximal rank (in the 3D case, of rank 3).  For a classical Hamiltonian system to be superintegrable, it must possess at least one additional functionally independent integral of motion. A 3D Hamiltonian system admitting a fourth integral of motion is called minimally superintegrable. If a 3D Hamiltonian system admits 5 integrals of motion, then such system is called maximally superintegrable.

It is known that the leading terms of an integral of motion, which is polynomial in momenta, take values in the enveloping algebra of the Euclidean algebra $\mathfrak{e}(3)$, i.e. a combination of the momentum components $p_i$ and of the angular momentum components $L_i=\epsilon_{ijk}x_jp_k$. Hence, a quadratic integral of motion takes the form
\begin{equation}
X=\sum_{1\leq i\leq j\leq 6}\alpha_{ij}Y^A_aY^A_b+\sum_{j=1}^3s_j(\vec{x})p^A_j+m(\vec{x}),\label{genx}
\end{equation} 
where
\begin{equation}
\hspace{-1cm}Y^A=(p^A_x,p^A_y,p^A_z,L^A_x,L^A_y,L^A_z),\qquad p^A_i=p_i+A_i(\vec{x}),\qquad L^A_i=\epsilon_{ijk}x_jp^A_k.\label{LAPA}
\end{equation}
The form (\ref{LAPA}) (including the vector potential components with the momenta) is convenient because the functions $s_j(\vec{x})$ and $m(\vec{x})$ together with the constants $\alpha_{ij}$ are gauge invariant. We recall that the Hamiltonian (\ref{genH}) and integrals (\ref{genx}) themselves are only gauge covariant. For a more detailed discussion, see e.g. \cite{MS18}, p. 2-3 and references there.

\section{Circular parabolic type integrability with magnetic fields}\label{SecInt}\setcounter{equation}{0}
In this section, we consider integrable Hamiltonian systems that admit two quadratic integrals of motion of the form
\begin{eqnarray}
X_1&=L_x^Ap_y^A-L_y^Ap_x^A+\mbox{lower order terms},\label{X1raw}\\
X_2&=(L_z^A)^2+\mbox{lower order terms}.\label{X2raw}
\end{eqnarray}
These integrals of motion correspond to the case where $\alpha_{24}=-\alpha_{15}=1$ in $X_1$, $\alpha_{66}=1$ in $X_2$ and all other $\alpha_{ij}$ are set to zero in the equation (\ref{genx}). Integrals of motion with such a structure imply the separation of variables in the Hamilton--Jacobi equation in the rotational parabolic coordinates  when the magnetic field vanishes \cite{MSVW,Eisenhart34}. The circular parabolic coordinates are given by the transformation
\begin{eqnarray}
x=\xi\eta\cos(\phi),\qquad y=\xi\eta\sin(\phi),\qquad z=\frac{1}{2}\left(\xi^2-\eta^2\right),\label{CPcoor}
\end{eqnarray}
where $\xi\in(0,\infty)$, $\eta\in(0,\infty)$ and $\phi\in(-\pi,\pi]$. (When $\xi=0$ or $\eta=0$, it corresponds to points on the $z$-axis where the rational parabolic coordinates fail to be single-valued.) The flat Cartesian metric $g_{ij}=\delta_{ij}$ in the new coordinates (\ref{CPcoor}) reads
\begin{equation}
g_{ij}=\left(\begin{array}{ccc}
\xi^2+\eta^2 & 0 & 0 \\
0 & \xi^2+\eta^2 & 0 \\
0 & 0 & \xi^2\eta^2
\end{array}\right).\label{metricCP}
\end{equation}
The pull-back of a 1-form, e.g.
\begin{equation}
A=A_xdx+A_ydy+A_zdz=A_\xi d\xi+A_\eta d\eta+A_\phi d\phi,
\end{equation}
is given by the relations
\begin{eqnarray}
A_x&=\frac{\cos (\phi )(\xi\,A_\eta+\eta\,A_\xi) }{\xi^2+\eta ^2}-\frac{\sin (\phi )A_\phi }{\xi \eta },\label{Axof}\nonumber\\
A_y&=\frac{\sin (\phi )(\xi\,A_\eta+\eta\,A_\xi) }{\xi^2+\eta ^2}+\frac{\cos (\phi )A_\phi }{\xi \eta },\label{Ayof}\\
A_z&=\frac{\xi\,A_\xi-\eta\,A_\eta}{\xi^2+\eta ^2}.\label{Azof}\nonumber
\end{eqnarray}
The transformation of the momentum $p$ satisfies the same equations as for a 1-form, c.f. equations (\ref{Ayof}).

In order to obtain the conditions for integrability and to solve the resulting partial differential equations for the integrals of the form (\ref{X1raw}) and (\ref{X2raw}), it is convenient to consider these equations in the circular parabolic coordinates. From (\ref{metricCP}), we find that the Hamiltonian takes the form
\begin{equation}
H=\frac{1}{2}\left(\frac{(p_\xi^A)^2}{\xi^2+\eta^2}+\frac{(p_\eta^A)^2}{\xi^2+\eta^2}+\frac{(p_\phi^A)^2}{\xi^2\eta^2}\right)+W(\xi,\eta,\phi)
\end{equation}
and the two integrals of motion become
\begin{eqnarray}
\hspace{-1cm}X_1=&\frac{\eta^2}{2(\xi^2+\eta^2)}(p_\xi^A)^2-\frac{\xi^2}{2(\xi^2+\eta^2)}(p_\eta^A)^2+\frac{1}{2}\left(\frac{1}{\xi^2}-\frac{1}{\eta^2}\right)(p_\phi^A)^2\nonumber\\
&+s^\xi_1(\xi,\eta,\phi)p^A_\xi+s^\eta_1(\xi,\eta,\phi)p^A_\eta+s^\phi_1(\xi,\eta,\phi)p^A_\phi+m_1(\xi,\eta,\phi),\label{CPX1gen}
\end{eqnarray}
\begin{equation}
\hspace{-1cm}X_2=(p_\phi^A)^2+s^\xi_2(\xi,\eta,\phi)p^A_\xi+s^\eta_2(\xi,\eta,\phi)p^A_\eta+s^\phi_2(\xi,\eta,\phi)p^A_\phi+m_2(\xi,\eta,\phi),\label{CPX2gen}
\end{equation}
where the functions $s^\xi_1$, $s^\xi_2$, $s^\eta_1$, $s^\eta_2$, $s^\phi_1$, $s^\phi_2$, $m_1$ and $m_2$ are to be determined together with the components of the magnetic field $B_\xi(\xi,\eta,\phi)$, $B_\eta(\xi,\eta,\phi)$, $B_\phi(\xi,\eta,\phi)$ and the scalar potential $W(\xi,\eta,\phi)$. The Poisson bracket in the circular parabolic coordinates becomes
\begin{equation}
\lbrace f,g\rbrace=\sum_{\alpha\in\lbrace\xi,\eta,\phi\rbrace}\frac{\partial f}{\partial\alpha}\frac{\partial g}{\partial p_\alpha}-\frac{\partial g}{\partial\alpha}\frac{\partial f}{\partial p_\alpha}.
\end{equation}
The conditions for integrability can be obtained by equating to zero the different coefficients of the polynomials in $p_\xi$, $p_\eta$ and $p_\phi$ of $\lbrace X_i,H\rbrace=0=\lbrace X_1,X_2\rbrace$. We are left with the second order equations 
\begin{eqnarray}
&\hspace{-2cm}\lbrace X_1,H\rbrace=0&\label{eq2order1}\\
&\hspace{-1.5cm}p_\xi^2:\qquad & \xi s^\xi_1+\eta s^\eta_1+\left(\xi^2+\eta ^2\right) \partial_\xi s^\xi_1=0,\nonumber\\
&\hspace{-1.5cm}p_\eta^2:\qquad & \xi s^\xi_1+\eta s^\eta_1+\left(\xi^2+\eta ^2\right) \partial_\eta s^\eta_1=0,\nonumber\\
&\hspace{-1.5cm}p_\phi^2:\qquad & \eta s^\xi_1+\xi s^\eta_1+\xi \eta  \partial_\phi s^\phi_1=0,\nonumber\\
&\hspace{-1.5cm}p_\xi p_\eta :\qquad & B_\phi=\partial_\eta s^\xi_1+\partial_\xi s^\eta_1,\nonumber\\
&\hspace{-1.5cm}p_\xi p_\phi :\qquad & B_\eta \xi^2+\left(\xi^2+\eta ^2\right) \partial_\phi s^\xi_1+\xi^2 \eta ^2 \partial_\xi s^\phi_1=0,\nonumber\\
&\hspace{-1.5cm}p_\eta p_\phi :\qquad & B_\xi \eta ^2+\left(\xi^2+\eta ^2\right) \partial_\phi s^\eta_1+\xi^2 \eta ^2 \partial_\eta s^\phi_1=0,\nonumber\\
&\hspace{-2cm}\lbrace X_2,H\rbrace=0&\label{eq2order2}\\
&\hspace{-1.5cm}p_\xi^2:\qquad & \left(\eta ^2+\xi^2\right) \partial_\xi s^\xi_2+\eta s^\eta_2+\xi s^\xi_2=0,\nonumber\\
&\hspace{-1.5cm}p_\eta^2:\qquad & \left(\eta ^2+\xi^2\right) \partial_\eta s^\eta_2+\eta s^\eta_2+\xi s^\xi_2=0,\nonumber\\
&\hspace{-1.5cm}p_\phi^2:\qquad & \xi s^\eta_2+\eta s^\xi_2+\eta  \xi \partial_\phi s^\phi_2=0,\nonumber\\
&\hspace{-1.5cm}p_\xi p_\eta :\qquad & \partial_\xi s^\eta_2+\partial_\eta s^\xi_2=0,\nonumber\\
&\hspace{-1.5cm}p_\xi p_\phi :\qquad & 2 \eta ^2 \xi^2 B_\eta=\left(\eta ^2+\xi^2\right) \partial_\phi s^\xi_2+\eta ^2 \xi^2 \partial_\xi s^\phi_2,\nonumber\\
&\hspace{-1.5cm}p_\eta p_\phi :\qquad & 2 \eta ^2 \xi^2 B_\xi+\left(\eta ^2+\xi^2\right) \partial_\phi s^\eta_2+\eta ^2 \xi^2 \partial_\eta s^\phi_2=0\nonumber,\\
&\hspace{-2cm}\lbrace X_1,X_2\rbrace=0&\qquad\mbox{(modulo equations (\ref{eq2order1})-(\ref{eq2order2}))}\label{eq2orderlast}\\
&\hspace{-1.5cm}p_\xi^2:\qquad & s^\eta_2=0,\nonumber\\
&\hspace{-1.5cm}p_\eta^2:\qquad & s^\xi_2=0,\nonumber\\
&\hspace{-1.5cm}p_\phi^2:\qquad & \xi \left(2 \eta  \xi s^\eta_1+2 \eta ^2 s^\xi_1+s^\xi_2\right)=\eta s^\eta_2,\nonumber\\
&\hspace{-1.5cm}p_\xi p_\eta :\qquad & \partial_\xi s^\eta_2=0,\nonumber\\
&\hspace{-1.5cm}p_\xi p_\phi :\qquad & 2 \eta ^2 \partial_\xi s^\phi_1+\partial_\xi s^\phi_2=0,\nonumber\\
&\hspace{-1.5cm}p_\eta p_\phi :\qquad & 2 \xi^2 \partial_\eta s^\phi_1=\partial_\eta s^\phi_2,\nonumber
\end{eqnarray}
and the lower order coefficients that we will consider further. It is straightforward to solve the system (\ref{eq2order1})-(\ref{eq2orderlast}) to get
\begin{eqnarray}
\hspace{-2cm}s^\xi_1(\xi,\eta ,\phi )=\frac{c_1\,\xi}{\xi^2+\eta^2},\quad s^\eta_1(\xi,\eta ,\phi )=\frac{-c_1\,\eta }{\xi^2+\eta^2},\quad s^\phi_1(\xi,\eta ,\phi )=\frac{f\left(\eta ^2\right)-g\left(\xi^2\right)}{\xi^2+\eta^2},\nonumber\\
\hspace{-1cm}s^\xi_2(\xi,\eta ,\phi )=s^\eta_2(\xi,\eta ,\phi )=0,\qquad s^\phi_2(\xi,\eta ,\phi )=\frac{2 \left(\xi^2 f\left(\eta ^2\right)+\eta ^2 g\left(\xi^2\right)\right)}{\eta ^2+\xi^2},
\end{eqnarray}
\begin{eqnarray}
B_\xi(\xi,\eta ,\phi )=\partial_\eta\left(\xi^2 \frac{g\left(\xi^2\right)-f\left(\eta ^2\right)}{\eta ^2+\xi^2}\right),\nonumber\\
B_\eta(\xi,\eta ,\phi )=\partial_\xi\left(\eta ^2\frac{ g\left(\xi^2\right)-f\left(\eta ^2\right)}{\eta ^2+\xi^2}\right),\qquad B_\phi(\xi,\eta ,\phi )=0.\label{solCP2}
\end{eqnarray}
where $c_1$ is an arbitrary constant and $f=f(\eta^2)$, $g=g(\xi^2)$ are arbitrary functions of $\eta^{2}$ and $\xi^{2}$, respectively.

The remaining first and zero order constrains read
\begin{eqnarray}
&\hspace{-2cm}\lbrace X_1,H\rbrace=0&\\
&\hspace{-1.5cm}p_\xi:\qquad2\eta^{2}\xi \left(f\left(\eta ^2\right)-g\left(\xi^2\right)\right) \left(f\left(\eta ^2\right)-g\left(\xi^2\right)+\left(\eta ^2+\xi^2\right) g'\left(\xi^2\right)\right)\nonumber\\
&+\left(\xi^2+\eta ^2\right)^3\left(\eta ^2 \partial_\xi W-\partial_\xi m_1\right) =0,\nonumber\\
&\hspace{-1.5cm}p_\eta:\qquad2\eta\xi^{2} \left(f\left(\eta ^2\right)-g\left(\xi^2\right)\right) \left(\left(\eta ^2+\xi^2\right) f'\left(\eta ^2\right)-f\left(\eta ^2\right)+g\left(\xi^2\right)\right)\nonumber\\
&-\left(\eta ^2+\xi^2\right)^3 \left(\partial_\eta m_1+\xi^2\partial_\eta W\right)=0,\nonumber\\
&\hspace{-1.5cm}p_\phi:\qquad -2c_1 \eta^{2}  \xi^{2} \left(\left(\eta ^2+\xi^2\right) \left(f'\left(\eta ^2\right)-g'\left(\xi^2\right)\right)-2 f\left(\eta ^2\right)+2 g\left(\xi^2\right)\right)\nonumber\\
&+\left(\eta ^2+\xi^2\right)^3 \left(\partial_\phi m_1+\left(\xi^2-\eta ^2\right)\partial_\phi W\right)=0,\nonumber\\
&\hspace{-1.5cm}1:\qquad c_1\left(\xi \partial_\xi W-\eta  \partial_\eta W\right)=\left(g\left(\xi^2\right)-f\left(\eta ^2\right)\right) \partial_\phi W,\nonumber\\
&\hspace{-2cm}\lbrace X_2,H\rbrace=0&\label{CPpphiX2}\\
&\hspace{-1.5cm}p_\xi:\qquad4 \eta ^2 \xi \left(\xi^2 f\left(\eta ^2\right)+\eta ^2 g\left(\xi^2\right)\right) \left(f\left(\eta ^2\right)+\left(\eta ^2+\xi^2\right) g'\left(\xi^2\right)-g\left(\xi^2\right)\right)\nonumber\\
&-\left(\eta ^2+\xi^2\right)^3 \partial_\xi m_2=0,\nonumber\\
&\hspace{-1.5cm}p_\eta:\qquad4 \eta  \xi^2 \left(\xi^2 f\left(\eta ^2\right)+\eta ^2 g\left(\xi^2\right)\right) \left(\left(\eta ^2+\xi^2\right) f'\left(\eta ^2\right)-f\left(\eta ^2\right)+g\left(\xi^2\right)\right)\nonumber\\
&-\left(\eta ^2+\xi^2\right)^3 \partial_\eta m_2=0,\nonumber\\
&\hspace{-1.5cm}p_\phi:\qquad\partial_\phi m_2=2 \eta ^2 \xi^2 \partial_\phi W,\nonumber\\
&\hspace{-1.5cm}1:\qquad\left(\xi^2 f\left(\eta ^2\right)+\eta ^2 g\left(\xi^2\right)\right) \partial_\phi W=0,\nonumber\\
&\hspace{-2cm}\lbrace X_1,X_2\rbrace=0&\\
&\hspace{-1.5cm}p_\phi:\qquad c_1\left(\left(\eta ^2+\xi^2\right) \left(f'\left(\eta ^2\right)-g'\left(\xi^2\right)\right)-2 f\left(\eta ^2\right)+2 g\left(\xi^2\right)\right)=0,\nonumber\\
&\hspace{-1.5cm}1:\qquad2 c_1 \eta ^2 \xi^2 \left(\xi^2 f\left(\eta ^2\right)+\eta ^2 g\left(\xi^2\right)\right) \left(\left(\eta ^2+\xi^2\right) \left(f'\left(\eta ^2\right)-g'\left(\xi^2\right)\right)-2 f\left(\eta ^2\right)+2 g\left(\xi^2\right)\right)\nonumber\\
&-\left(\eta ^2+\xi^2\right)^3 \left(\xi^4 f\left(\eta ^2\right)-\eta ^4 g\left(\xi^2\right)\right) \partial_\phi W=0.\nonumber
\end{eqnarray}
From the coefficient $1$ of equation (\ref{CPpphiX2}), one can see that either the potential $W$ does not depend on $\phi$ or that $f(\eta^2)=\lambda\eta^2$ and $g(\xi^2)=-\lambda\xi^2$. However, the last case leads to a vanishing magnetic field, which has already been studied in \cite{MSVW}. Hence, we will only consider the case where $\partial_\phi W=0$. By solving every equation which does not involve $c_1$, we obtain that
\begin{eqnarray}
&m_1(\xi,\eta,\phi)=\frac{\xi^4 \alpha \left(\eta ^2\right)+\eta ^4 \beta \left(\xi^2\right)}{\eta ^2 \xi^2 \left(\eta ^2+\xi^2\right)},\nonumber\\
&m_2(\xi,\eta,\phi)=\left(\frac{\xi^2 f\left(\eta ^2\right)+\eta ^2 g\left(\xi^2\right)}{\eta ^2+\xi^2}\right)^2,
\end{eqnarray}
\begin{eqnarray}
&W(\xi,\eta,\phi)=\frac{1}{2}\left(\frac{f\left(\eta ^2\right)-g\left(\xi^2\right)}{\eta ^2+\xi^2}\right)^2+\frac{\eta^2\beta \left(\xi^2\right)-\xi^2\alpha \left(\eta ^2\right)}{\xi^2\eta^2\left(\eta ^2+\xi^2\right)},\label{solCP1}
\end{eqnarray}
where $\alpha(\eta^2)$ and $\beta(\xi^2)$ are arbitrary functions of $\eta^{2}$ and $\xi^{2}$ respectively. We are still left with two equations, i.e.
\begin{equation}
c_1\left(\left(\eta ^2+\xi^2\right) \left(f'\left(\eta ^2\right)-g'\left(\xi^2\right)\right)-2 f\left(\eta ^2\right)+2 g\left(\xi^2\right)\right)=0\label{remains1}
\end{equation}
and
\begin{equation}
c_1\left(\xi \partial_\xi W-\eta  \partial_\eta W\right)=0.\label{remains2}
\end{equation}
Thus, we either have that $c_1=0$ and the integrals are, up to a shift by a constant and the transformation (\ref{DELTA}) below, fully fixed by the magnetic field (\ref{solCP2}) and the potential (\ref{solCP1}) or that $f$, $g$ and $W$ satisfy equations (\ref{remains1}) and (\ref{remains2}) for any value of $c_1$. In that case we have an additional first order integral, i.e. we are in an superintegrable subcase that we will discuss in the section \ref{SecSup}.

As a result, we get that all integrable Hamiltonian systems that admit the pair of integrals of motion (\ref{X1raw}) and (\ref{X2raw}) can be described by the following Hamiltonian,
\begin{equation}
\hspace{-2.5cm}H=\frac{1}{2}\left(\frac{(p_\xi^A)^2}{\xi^2+\eta^2}+\frac{(p_\eta^A)^2}{\xi^2+\eta^2}+\frac{(p_\phi^A)^2}{\xi^2\eta^2}\right)+\frac{1}{2}\left(\frac{f\left(\eta ^2\right)-g\left(\xi^2\right)}{\eta ^2+\xi^2}\right)^2+\frac{\eta^2\beta \left(\xi^2\right)-\xi^2\alpha \left(\eta ^2\right)}{\xi^2\eta^2\left(\eta ^2+\xi^2\right)},
\end{equation}
where 4 arbitrary functions $\alpha(\eta^2)$, $\beta(\xi^2)$, $f(\eta^2)$ and $g(\xi^2)$ appear. The associated magnetic field involves two arbitrary functions $f(\eta^2)$ and $g(\xi^2)$, i.e.
\begin{eqnarray}
B_\xi=&-2 \eta  \xi^2 \frac{\left(\eta ^2+\xi^2\right) f'\left(\eta ^2\right)-f\left(\eta ^2\right)+g\left(\xi^2\right)}{\left(\eta ^2+\xi^2\right)^2},\nonumber\\
B_\eta=&2 \eta ^2 \xi\frac{ \left(\eta ^2+\xi^2\right) g'\left(\xi^2\right)+f\left(\eta ^2\right)-g\left(\xi^2\right)}{\left(\eta ^2+\xi^2\right)^2},\\
B_\phi=&0.\nonumber
\end{eqnarray}
The vector potential can be chosen as
\begin{equation}
A_\xi=A_\eta=0,\qquad A_\phi(\xi,\eta)=-\frac{\xi^2 f\left(\eta ^2\right)+\eta ^2 g\left(\xi^2\right)}{\eta ^2+\xi^2},\label{CPintA}
\end{equation}
up to a gauge transformation.

It is interesting to observe that the magnetic field is not at all constrained by the first and zeroth order determining equations. In all the previously investigated, subgroup type, cases, the magnetic field was always somehow constrained by the lower order determining equations, which is not the case here. As we will see later, all three non-subgroup types considered here possess this feature. We don't know yet whether it is a property of non-subgroup type classes in general or that it is a consequence of the fact that for these three classes one quadratic integral of motion always reduces to a first order integral.

The two integrals of motion (\ref{X1raw}) and (\ref{X2raw}) take the explicit forms
\begin{eqnarray}
X_1=&\frac{\eta ^2 (p_\xi^A)^2-\xi^2(p_\eta^A)^2 }{2 \left(\eta ^2+\xi^2\right)}+\frac{1}{2}  \left(\frac{1}{\xi^2}-\frac{1}{\eta ^2}\right)(p_\phi^A)^2\nonumber\\
&+ \left(\frac{f\left(\eta ^2\right)-g\left(\xi^2\right)}{\eta ^2+\xi^2}\right)p_\phi^A+\frac{\xi^4 \alpha \left(\eta ^2\right)+\eta ^4 \beta \left(\xi^2\right)}{\eta^2\xi^2(\eta ^2 + \xi^2)},\\
X_2=\, &\left(p_\phi^A+\frac{\xi^2 f\left(\eta ^2\right)+\eta ^2 g\left(\xi^2\right)}{\eta ^2+\xi^2}\right)^2.
\end{eqnarray}
That means that $X_2$ is equivalent to a first order integral of motion, i.e.
\begin{equation}
\tilde{X}_2=\sqrt{X_2}=p_\phi^{A}+\frac{\xi^2 f\left(\eta ^2\right)+\eta ^2 g\left(\xi^2\right)}{\eta ^2+\xi^2},
\end{equation}
which is equal to $L_z$ with the choice of the vector potential (\ref{CPintA}). The consistency of our solution requires that $X_1$ possesses a freedom of adding $\tilde{X}_2$
to it without changing the magnetic field or the potential. That can be indeed accomplished by an alternative choice
\begin{equation}
\hat{f}= f+\lambda\eta^2,\quad \hat{g}= g-\lambda\xi^2,\quad \hat{\alpha}=\alpha+\lambda\eta^2f,\quad \hat{\beta}=\beta+\lambda\xi^2g\label{DELTA}
\end{equation}
of our arbitrary functions, which leaves $\vec{B}$ and $W$ the same.

\section{Oblate spheroidal type integrability with magnetic fields}\label{SecOblate}\setcounter{equation}{0}
In this section, we consider integrable Hamiltonian systems that admit two quadratic integrals of motion of the form
\begin{eqnarray}
X_1&=(L^A)^2+a^2((p^A_x)^2+(p^A_y)^2)+\mbox{lower order terms},\label{X1Oraw}\\
X_2&=(L_z^A)^2+\mbox{lower order terms}.\label{X2Oraw}
\end{eqnarray}
These integrals of motion correspond to the case where $\alpha_{11}=\alpha_{22}=a^2$ together with $\alpha_{44}=\alpha_{55}=\alpha_{66}=1$ in $X_1$, $\alpha_{66}=1$ in $X_2$ and all other $\alpha_{ij}$ are set to zero in the equation (\ref{genx}). Integrals of motion with such a structure imply the separation of variables in the oblate spheroidal coordinates in the Hamilton--Jacobi equation when the magnetic field vanishes \cite{MSVW,Eisenhart34}. The oblate spheroidal coordinates are given by the transformation
\begin{eqnarray}
\hspace{-2cm}x=a\cosh(\xi)\sin(\eta)\cos(\phi),\quad
y=a\cosh(\xi)\sin(\eta)\sin(\phi),\quad
z=a\sinh(\xi)\cos(\eta),
\end{eqnarray}
where $\xi\in(0,\infty)$, $\eta\in(0,\pi)$, $\phi\in(-\pi,\pi]$ and the parameter $a>0$. (When $\xi=0$ or $\eta=0,\pi$, the coordinates are multiple-valued --- again this corresponds to points along the $z$-axis.) The metric in the oblate spheroidal coordinates takes the form
\begin{equation}
\hspace{-2cm}g_{ij}=\left(\begin{array}{ccc}
a^2(\cosh^2(\xi)-\sin^2(\eta)) & 0 & 0 \\
0 & a^2(\cosh^2(\xi)-\sin^2(\eta)) & 0 \\
0 & 0 & a^2\cosh^2(\xi)\sin^2(\eta)
\end{array}\right)
\end{equation}
and the pull-back of a 1-form, e.g. of the vector potential 1-form,
\begin{equation}
A=A_xdx+A_ydy+A_zdz=A_\xi d\xi+A_\eta d\eta+A_\phi d\phi
\end{equation}
is given by the relations
\begin{eqnarray}
\hspace{-1cm}A_x&=\frac{ \cos (\phi ) (A_\eta\cos (\eta ) \cosh (\xi)+A_\xi \sin (\eta ) \sinh (\xi))}{a(\cosh^2 ( \xi)-\sin^2 ( \eta ))}-\frac{A_\phi\sin (\phi )}{a\cosh(\xi)\sin(\eta)},\nonumber\\
\hspace{-1cm}A_y&=\frac{ \sin (\phi ) (A_\eta\cos (\eta ) \cosh (\xi)+A_\xi \sin (\eta ) \sinh (\xi))}{a(\cosh^2 ( \xi)-\sin^2 ( \eta ))}+\frac{A_\phi\cos (\phi )}{a\cosh(\xi)\sin(\eta)},\\
\hspace{-1cm}A_z&=\frac{ A_\xi \cos (\eta ) \cosh (\xi)-A_\eta\sin (\eta ) \sinh (\xi)}{a (\cosh^2 (\xi)-\cos^2 ( \eta ))}.\nonumber
\end{eqnarray}

The determining equations can be solved in a very similar way as in the circular parabolic case. 
The second order determining equations prescribe the structure of the magnetic field and of the terms linear in momenta in (\ref{X1Oraw}) and (\ref{X2Oraw}).
We also find that if the scalar potential depends on $\phi$ then the magnetic field vanishes. There is again a constant that appears in the solution of the second order equations, which is associated with an additional first order integral of motion, i.e. a superintegrable subcase, that is equivalent to the case \ref{Scase1} arising in the circular parabolic case, see below.

As a result, in the oblate spheroidal coordinates, the Hamiltonian is
\begin{eqnarray}
\hspace{-1cm}H=&\frac{1}{2}\left(\frac{(p_\xi^A)^2+(p_\eta^A)^2}{a^2(\cosh^2(\xi)-\sin^2(\eta))}+\frac{(p_\phi^A)^2}{a^2\cosh^2(\xi)\sin^2(\eta)}\right)\nonumber\\
\hspace{-1cm}&+\frac{\alpha(\eta)+\beta(\xi)}{2a^2(\cosh^2(\xi)-\sin^2(\eta))}-\frac{1}{2}\left(\frac{f(\eta)-g(\xi)}{2a(\cosh^2(\xi)-\sin^2(\eta))}\right)^2
\end{eqnarray}
where 4 arbitrary functions $\alpha(\eta)$, $\beta(\xi)$, $f(\eta)$ and $g(\xi)$ appear, and the vector potential can be chosen as
\begin{equation}
A_\xi=A_\eta=0,\qquad A_\phi(\xi,\eta)=\frac{\sin^2(\eta)g(\xi)-\cosh^2(\xi)f(\eta)}{2(\cosh^2(\xi)-\sin^2(\eta))},\label{CPA}
\end{equation}
up to a gauge transformation. The associated magnetic field involves two arbitrary functions $f(\eta)$ and $g(\xi)$, i.e.
\begin{eqnarray}
\hspace{-2cm}B_\xi=&\partial_\eta A_\phi=\frac{(g(\xi)-f(\eta))\sin(\eta)\cos(\eta)\cosh^2(\xi)}{(\cosh^2(\xi)-\sin^2(\eta))^2}-\frac{
\cosh^2(\xi)f'(\eta)}{2(\cosh^2(\xi)-\sin^2(\eta))},\nonumber\\
\hspace{-2cm}B_\eta=&-\partial_\xi A_\phi=\frac{(g(\xi)-f(\eta))\sin^2(\eta)\sinh(\xi)\cosh(\xi)}{(\cosh^2(\xi)-\sin^2(\eta))^2}-\frac{
\sin^2(\eta)g'(\xi)}{2(\cosh^2(\xi)-\sin^2(\eta))},\\
\hspace{-2cm}B_\phi=&0.\nonumber
\end{eqnarray}
Again, it is interesting to note that the magnetic field is not at all constrained by the first and zeroth order constraints, like in the circular parabolic case. The two integrals of motion (\ref{X1raw}) and (\ref{X2raw}) take the explicit forms
\begin{eqnarray}
X_1=&\frac{\sin^2(\eta)(p_\xi^A)^2+\cosh^2(\xi)(p_\eta^A)^2}{\cosh^2(\xi)-\sin^2(\eta)}+\frac{\cosh^2(\xi)+\sin^2(\eta)}{\cosh^2(\xi)\sin^2(\eta)}(p_\phi^A)^2\nonumber\\
&+\left(\frac{f(\eta)-g(\xi)}{\cosh^2(\xi)-\sin^2(\eta)}\right)p_\phi^A+\frac{\cosh^2(\xi)\alpha(\eta)+\sin^2(\eta)\beta(\xi)}{\cosh^2(\xi)-\sin^2(\eta)},\\
X_2=&\left(p_\phi^A+\frac{\cosh^2(\xi)f(\eta)-\sin^2(\eta)g(\xi)}{2(\cosh^2(\xi)-\sin^2(\eta))}\right)^2,
\end{eqnarray}
where $X_2$ is equivalent to a first order integral of motion
\begin{equation}
\tilde{X}_2=p_\phi^A+\frac{\cosh^2(\xi)f(\eta)-\sin^2(\eta)g(\xi)}{2(\cosh^2(\xi)-\sin^2(\eta))},
\end{equation}
which in our choice of gauge reduces to $L_z$. 

The superintegrable subcase \ref{Scase1} is obtained when
\begin{eqnarray}
f(\eta )=a^2 b_z \sin ^4(\eta ),\qquad g(\xi)= a^2 b_z \cosh ^4(\xi),\nonumber\\
\alpha (\eta )= \frac{5a^3 b_z^2}{64}-\frac{a^3 b_z^2}{4}\sin^6(\eta)+\frac{2 \omega  }{a\sin ^2(\eta )},\\
\beta (\xi)=-\frac{5a^3 b_z^2}{64}+\frac{a^3 b_z^2}{4}\cosh^6(\xi)-\frac{2 \omega  }{a\cosh^2(\xi)}.\nonumber
\end{eqnarray}
When it is expressed in Cartesian coordinates, we find that it is identical to the case \ref{Scase1}, treated in section \ref{SecSup} below.

\section{Prolate spheroidal type integrability with magnetic fields}\label{SecProlate}\setcounter{equation}{0}
Next, we consider integrable Hamiltonian systems that admit two quadratic integrals of motion of the form
\begin{eqnarray}
X_1&=(L^A)^2-a^2((p^A_x)^2+(p^A_y)^2)+\mbox{lower order terms},\label{X1Praw}\\
X_2&=(L_z^A)^2+\mbox{lower order terms}.\label{X2Praw}
\end{eqnarray}
These integrals of motion correspond to the case where $\alpha_{11}=\alpha_{22}=-a^2$ together with $\alpha_{44}=\alpha_{55}=\alpha_{66}=1$ in $X_1$, $\alpha_{66}=1$ in $X_2$ and all other $\alpha_{ij}$ are set to zero in the equation (\ref{genx}), i.e. they differ from the oblate spheroidal type by the sign of the $a^2$ term. Integrals of motion with such a structure imply the separation of variables in the prolate spheroidal coordinates of the Hamilton--Jacobi equations when the magnetic field vanishes \cite{MSVW}. The prolate spheroidal coordinates are given by the transformation
\begin{eqnarray}
\hspace{-2cm}x=a\sinh(\xi)\sin(\eta)\cos(\phi),\quad y=a\sinh(\xi)\sin(\eta)\sin(\phi),\quad z=a\cosh(\xi)\cos(\eta),
\end{eqnarray}
where $\xi\in(0,\infty)$, $\eta\in(0,\pi)$, $\phi\in(-\pi,\pi]$ and the parameter $a>0$. (When $\xi=0$ or $\eta=0,\pi$, the coordinates are multiple-valued, again along the $z$-axis.) They differ from the oblate spheroidal coordinates by the interchange of $\sinh(\xi)$ and $\cosh(\xi)$. The new metric takes the form
\begin{equation}
\hspace{-2cm}g_{ij}=\left(\begin{array}{ccc}
a^2(\sinh^2(\xi)+\sin^2(\eta)) & 0 & 0 \\
0 & a^2(\sinh^2(\xi)+\sin^2(\eta)) & 0 \\
0 & 0 & a^2\sinh^2(\xi)\sin^2(\eta)
\end{array}\right).
\end{equation}
and the pull-back of a 1-form, e.g. for
\begin{equation}
A=A_xdx+A_ydy+A_zdz=A_\xi d\xi+A_\eta d\eta+A_\phi d\phi
\end{equation}
is given by the relations
\begin{eqnarray}
\hspace{-1cm}A_x&=\frac{ \cos (\phi ) (A_\eta\cos (\eta ) \sinh (\xi)+A_\xi \sin (\eta ) \cosh (\xi))}{a(\sinh^2(\xi)+\sin^2(\eta))}-\frac{A_\phi\sin (\phi )}{a\sinh(\xi)\sin(\eta)},\nonumber\\
\hspace{-1cm}A_y&=\frac{ \sin (\phi ) (A_\eta\cos (\eta ) \sinh (\xi)+A_\xi \sin (\eta ) \cosh (\xi))}{a(\sinh^2(\xi)+\sin^2(\eta))}+\frac{A_\phi\cos (\phi )}{a\sinh(\xi)\sin(\eta)},\\
\hspace{-1cm}A_z&=\frac{ A_\xi \cos (\eta ) \sinh (\xi)-A_\eta\sin (\eta ) \cosh (\xi)}{a (\sinh^2(\xi)+\sin^2(\eta))}.\nonumber
\end{eqnarray}
The determining equations can be solved in a very similar way as in the circular parabolic case (and the oblate spheroidal case). The second order determining equations prescribe the structure of the magnetic field and of the terms linear in momenta in (\ref{X1Praw}) and (\ref{X2Praw}). We find again that if the scalar potential depends on $\phi$ then the magnetic field vanishes. There is also a constant that appears in the solution of the second order equations, which is associated with an additional first order integral of motion, i.e. a superintegrable subcase, that is equivalent to the case \ref{Scase1} arising in both previous cases.

As a result, in the prolate spheroidal coordinates, the Hamiltonian is
\begin{eqnarray}
\hspace{-1cm}H=&\frac{1}{2}\left(\frac{(p_\xi^A)^2+(p_\eta^A)^2}{a^2(\sinh^2(\xi)+\sin^2(\eta))}+\frac{(p_\phi^A)^2}{a^2\sinh^2(\xi)\sin^2(\eta)}\right)\nonumber\\
\hspace{-1cm}&+\frac{\alpha(\eta)+\beta(\xi)}{2a^2(\sinh^2(\xi)+\sin^2(\eta))}+\frac{1}{8}\left(\frac{f(\eta)-g(\xi)}{a(\sinh^2(\xi)+\sin^2(\eta))}\right)^2
\end{eqnarray}
where 4 arbitrary functions $\alpha(\eta)$, $\beta(\xi)$, $f(\eta)$ and $g(\xi)$ appear, and the vector potential can be chosen as
\begin{equation}
A_\xi=A_\eta=0,\qquad A_\phi(\xi,\eta)=-\frac{\sin^2(\eta)g(\xi)+\sinh^2(\xi)f(\eta)}{2(\sinh^2(\xi)+\sin^2(\eta))},\label{APS}
\end{equation}
up to a gauge transformation. The associated magnetic field involves two arbitrary functions $f(\eta)$ and $g(\xi)$, i.e.
\begin{eqnarray}
\hspace{-2cm}B_\xi=&\partial_\eta A_\phi=\frac{(f(\eta)-g(\xi))\sin(\eta)\cos(\eta)\sinh^2(\xi)}{(\sinh^2(\xi)+\sin^2(\eta))^2}-\frac{
\sinh^2(\xi)f'(\eta)}{2(\sinh^2(\xi)+\sin^2(\eta))},\nonumber\\
\hspace{-2cm}B_\eta=&-\partial_\xi A_\phi=\frac{(f(\eta)-g(\xi))\sin^2(\eta)\sinh(\xi)\cosh(\xi)}{(\sinh^2(\xi)+\sin^2(\eta))^2}+\frac{
\sin^2(\eta)g'(\xi)}{2(\sinh^2(\xi)+\sin^2(\eta))},\\
\hspace{-2cm}B_\phi=&0.\nonumber
\end{eqnarray}
Again, it is interesting to note that the magnetic field is not at all constrained by the first and zero order constraints, as in the two previous cases. The two integrals of motion (\ref{X1raw}) and (\ref{X2raw}) take the explicit forms
\begin{eqnarray}
X_1=&\frac{\sinh^2(\xi)(p_\eta^A)^2-\sin^2(\eta)(p_\xi^A)^2}{\sinh^2(\xi)+\sin^2(\eta)}+\frac{\sinh^2(\xi)-\sin^2(\eta)}{\sinh^2(\xi)\sin^2(\eta)}(p_\phi^A)^2\nonumber\\
&+\left(\frac{g(\xi)-f(\eta)}{\sinh^2(\xi)+\sin^2(\eta)}\right)p_\phi^A+\frac{\sinh^2(\xi)\alpha(\eta)-\sin^2(\eta)\beta(\xi)}{\sinh^2(\xi)+\sin^2(\eta)},\\
X_2=&\left(p_\phi^A+\frac{\sinh^2(\xi)f(\eta)+\sin^2(\eta)g(\xi)}{2(\sinh^2(\xi)+\sin^2(\eta))}\right)^2,
\end{eqnarray}
where $X_2$ is equivalent to a first order integral of motion, i.e.
\begin{equation}
\tilde{X}_2=p_\phi^A+\frac{\sinh^2(\xi)f(\eta)+\sin^2(\eta)g(\xi)}{2(\sinh^2(\xi)+\sin^2(\eta))},
\end{equation}
which reduces to $L_z$ in our gauge choice (\ref{APS}). 

The superintegrable subcase is obtained when 
\begin{eqnarray}
f(\eta )= -a^2 b_z \sin ^4(\eta ),\qquad g(\xi)= -a^2 b_z \sinh ^4(\xi),\nonumber\\
\alpha (\eta )= \frac{5a^3 b_z^2}{64}-\frac{a^3 b_z^2}{4}\sin^6(\eta)+\frac{2 \omega  }{a\sin ^2(\eta )},\\
\beta (\xi)=-\frac{5a^3 b_z^2}{64}-\frac{a^3 b_z^2}{4}\sinh^6(\xi)+\frac{2 \omega  }{a\sinh^2(\xi)}.\nonumber
\end{eqnarray}
When it is expressed in Cartesian coordinates, we find that it is identical to the case \ref{Scase1}.

\section{First-order additional integrals of motion for superintegrability}\label{SecSup}\setcounter{equation}{0}
In this section, for the circular parabolic case, we are looking in a systematic way for additional integrals of motion which are of first order in momenta, i.e.
\begin{equation}
\hspace{-1cm}Y=k_{1}p^A_x+k_{2}p^A_y+k_3p^A_z+k_{4}L^A_x+k_{5}L^A_y+m_3(x,y,z).\label{firstY}
\end{equation}
We have excluded the dependency in $L_z^A$ since the new integral of motion $Y$ can be modified by an addition of $\tilde{X_2}$ without any loss of generality. In the circular parabolic coordinates, the first order coefficients of (\ref{firstY}) can be written as
\begin{eqnarray}
\hspace{-2cm}s^\xi_3(\xi,\eta ,\phi )=&\frac{ \eta \cos (\phi ) \left(2k_1+k_5 \left(\eta ^2+\xi^2\right)\right)+ \theta  \sin (\phi ) \left(2k_2+k_4 \left(\eta ^2+\xi^2\right)\right)+2k_3 \xi}{ 2\left(\eta ^2+\xi^2\right)},\nonumber\\
\hspace{-2cm}s^\eta_3(\xi,\eta ,\phi )=&\frac{ \xi \cos (\phi ) \left(2k_1-k_5 \left(\eta ^2+\xi^2\right)\right)+ \xi \sin (\phi ) \left(2k_2-k_4 \left(\eta ^2+\xi^2\right)\right)-  2k_3\eta}{2 \left(\eta ^2+\xi^2\right)},\\
\hspace{-2cm}s^\phi_3(\xi,\eta ,\phi )=&\frac{\cos (\phi ) \left(2k_2+k_4 \left(\eta ^2-\xi^2\right)\right)-\sin (\phi ) \left(2k_1+k_5 \left(\eta ^2-\xi^2\right)\right)}{2\eta \xi}.\nonumber
\end{eqnarray}
The Poisson bracket of $Y$ with the Hamiltonian involves only terms linear and constant in momenta. The first order determining equations can be solved with respect to the derivatives $\partial_\xi m_3$, $\partial_\eta m_3$ and $\partial_\phi m_3$ of the function $m_3$ and do not depend on $m_3$ itself. These, in turn, imply compatibility conditions, e.g.
\begin{equation}
\partial_\xi\left(\partial_\eta m_3\right)=\partial_\eta\left(\partial_{\xi} m_3\right),
\end{equation}
which involve only the constants $k_1$, ..., $k_5$ and the functions $f$, $g$, $\alpha$ and $\beta$. Solving these equations, we arrive (under the assumption that the magnetic field doesn't vanish) at only three possibilities, namely
\begin{enumerate}
\item $k_3\neq0$ and $k_1=k_2=k_4=k_5=0$,\label{Scase1}
\item $k_1\neq0$ and $k_3=k_4=k_5=0$,\label{Scase2}
\item $k_4\neq0$ and $k_1=k_2=k_3=0$.\label{Scase3}
\end{enumerate}
In the cases \ref{Scase2} and \ref{Scase3}, there are no constraints on $k_2$ and $k_5$, respectively. Hence, we obtain in both cases a fifth independent integral of motion.

\subsection{Case \ref{Scase1}, $k_1=k_2=k_4=k_5=0.$}
This case is equivalent to the case where $c_1\neq0$, in (\ref{remains1}) and (\ref{remains2}) appearing as a subcase in the integrable case, where $4k_3=c_1$. The remaining constraints on $f(\eta^2)$, $g(\xi^2)$, $\alpha(\eta^2)$ and $\beta(\xi^2)$ can be solved and reduced (by adding suitable constant terms to $H$, $X_1$ and $\tilde{X_2}$) to
\begin{eqnarray}
\hspace{-2cm}f(\eta^2)=-\frac{b_z}{2}\eta^4,\quad g(\xi^2)=-\frac{b_z}{2}\xi^4,\quad \alpha(\eta^2)=-\omega+\frac{b_z^2}{8}\eta^8,\quad \beta(\xi^2)=\omega-\frac{b_z^2}{8}\xi^8.
\end{eqnarray}
Hence, the magnetic field takes the form
\begin{equation}
B_\xi=b_z \eta  \xi^2,\qquad B_\eta=-b_z \eta^2  \xi,\qquad B_\phi=0
\end{equation}
with a convenient choice of the vector potential as
\begin{equation}
A_\xi=A_\eta=0\qquad A_\phi=\frac{1}{2} b_z \xi^2 \eta ^2.
\end{equation}
The associated Hamiltonian is
\begin{equation}
H=\frac{1}{2}\left(\frac{(p_\xi^A)^2+(p_\eta^A)^2}{\left(\xi^2+\eta ^2\right)}+\frac{(p_\phi^A)^2}{\xi^2 \eta ^2}\right)+\frac{\omega}{\eta ^2 \xi^2}-\frac{b_z^2}{8}  \eta ^2 \xi^2
\end{equation}
and the integrals of motion are given by
\begin{eqnarray}
X_1=\frac{\eta ^2 (p_\xi^A)^2- \xi^2(p_\theta^A)^2}{2 \left(\eta ^2+\xi^2\right)}-\frac{\xi^2-\eta^2}{2\xi^2\eta^2}\left(\left(p_\phi^A-\frac{1}{2} b_z \xi^2 \eta ^2\right)^2+2\omega\right),\nonumber\\
\tilde{X_2}=p_\phi^A-\frac{1}{2} b_z \xi^2 \eta ^2,\\
Y_3=\frac{p_\xi^A \xi-\eta p_\eta^A}{\eta ^2+\xi^2}.\nonumber
\end{eqnarray}
Pulling back these results into the Cartesian coordinates, we find a constant magnetic field in the direction of the $z$-axis, i.e.
\begin{eqnarray}
(B_x,B_y,B_z)=(0,0,b_z),\qquad (A_x,A_y,A_z)=\left(-\frac{b_zy}{2},\frac{b_zx}{2},0\right),
\end{eqnarray}
and the integrals of motion read
\begin{eqnarray}
H=\frac{1}{2}\left((p_x^A)^2+(p_y^A)^2+(p_z^A)^2\right)-\frac{1}{8} b_z^2 \left(x^2+y^2\right)+\frac{\omega}{x^2+y^2},\nonumber\\
X_1=L_x^Ap_y^A-L_y^Ap_x^A+b_z zL_z^A-\frac{1}{4} b_z^2 z \left(x^2+y^2\right)-\frac{2 \omega z}{x^2+y^2},\label{equation}\\
\tilde{X}_2=L_z^A-\frac{1}{2} b_z \left(x^2+y^2\right),\qquad Y_3=p_z^A,\nonumber
\end{eqnarray}
Such a system appears in \cite{MSW18}, see class III therein, and is characterized by free motion along the $z$-axis together with a constant magnetic field oriented along the $z$-axis. Also in \cite{MSW18} (p.16), all the integrals of this system that are at most quadratic in momenta were found. In addition to the integrals (\ref{equation}), there is another functionally dependent quadratic integral with the leading order term $(L^A)^2$. This makes immediately obvious that this system belongs to the intersection of all three classes of integrable systems considered here.

\subsection{Case \ref{Scase2}, $k_3=k_4=k_5=0.$}
For this case, the remaining constraints on $f(\eta^2)$, $g(\xi^2)$, $\alpha(\eta^2)$ and $\beta(\xi^2)$ can be solved and reduced by constant shifts of $H$, $X_1$ and $X_2$ and a translation in $z$ to
\begin{equation}
f(\eta^2)=-\frac{b_z}{2}\eta^4,\quad g(\xi^2)=-\frac{b_z}{2}\xi^4,\quad\alpha(\eta^2)=\beta(\xi^2)=0.
\end{equation}
The magnetic field and its vector potential are identical to the preceding case, i.e.
\begin{equation}
\hspace{-1cm}(B_\xi,B_\eta,B_\phi)=(b_z \eta  \xi^2,-b_z \eta^2  \xi,0),\qquad (A_\xi,A_\eta,A_\phi)=\left(0,0,\frac{b_z}{2}  \xi^2 \eta ^2\right).
\end{equation}
However, the Hamiltonian and the integrals of motion are different, i.e.
\begin{eqnarray}
H=\frac{1}{2}\left(\frac{(p_\xi^A)^2+(p_\eta^A)^2}{ \left(\eta ^2+\xi^2\right)}+\frac{(p_\phi^A)^2}{ \eta ^2 \xi^2}\right)+\frac{b_z^2}{8} \left(\xi^2-\eta ^2\right)^2,\nonumber\\
X_1=\frac{\eta ^2 (p_\xi^A)^2- \xi^2(p_\eta^A)^2}{2 \left(\eta ^2+\xi^2\right)}+\frac{ \left(\eta^2-\xi ^2\right)}{2 \eta ^2 \xi^2}(p_\phi^A)^2+\frac{b_z}{2}\left(\xi ^2-\eta^2\right)p_\phi^A, \nonumber\\
\tilde{X}_2=p_\phi^A-\frac{b_z}{2}  \xi^2 \eta ^2,\\
Y_3=\frac{p_\eta^A \xi+p_\xi^A \eta }{\xi^2+\eta ^2}\cos (\phi )+\left(-\frac{p_\phi^A }{\xi \eta }+b_z \xi \eta\right)  \sin (\phi ),\nonumber\\
Y_4=\frac{p_\eta^A \xi+p_\xi^A \eta }{\xi^2+\eta ^2}\sin (\phi )+\left(\frac{p_\phi^A }{\xi \eta }-b_z \xi \eta\right)  \cos (\phi ),\nonumber
\end{eqnarray}
thus we get a maximally superintegrable Hamiltonian system. In the Cartesian coordinates, it reads
\begin{eqnarray}
(B_x,B_y,B_z)=(0,0,b_z),\qquad (A_x,A_y,A_z)=\left(-\frac{b_z y}{2},\frac{b_z x}{2},0\right),\nonumber\\
H=\frac{1}{2}\left((p_x^A)^2+(p_y^A)^2+(p_z^A)^2\right)+\frac{b_z^2 z^2}{2},\nonumber\\
X_1=L_x^Ap_y^A-L_y^Ap_x^A+b_z  zL_z^A,\\
\tilde{X}_2=L_z^A-\frac{1}{2} b_z \left(x^2+y^2\right),\nonumber\\
Y_3=p_x^A+b_z y,\qquad Y_4=p_y^A-b_z x\nonumber
\end{eqnarray}
Such a system appears in \cite{MS17}, p. 17 and is characterized by a constant magnetic field along the $z$-axis, but is not equivalent to the previous case \ref{Scase1} due to $z$-dependence of the scalar potential $W$. In addition, this system can also be  interpreted as the center of mass component of the Hamiltonian of the two-electron quantum dots for certain special values of the magnetic field and of the confinement frequencies, see e.g. \cite{BNY13,ZZHG}.

\subsection{Case \ref{Scase3}, $k_1=k_2=k_3=0$.}
By solving the remaining equations, we find after getting rid of irrelevant constants that
\begin{equation}
f(\eta^2)=b_m,\qquad g(\xi^2)=-b_m,\qquad \alpha(\eta^2)=0,\qquad \beta(\xi^2)=\omega.
\end{equation}
Hence, the magnetic field and the vector potential become
\begin{equation}
\hspace{-2.5cm}\left(B_\xi,B_\eta,B_\phi\right)=\left(\frac{4 b_m \xi^2 \eta}{\left(\xi^2+\eta ^2\right)^2},\frac{4 b_m \xi \eta^2}{\left(\xi^2+\eta ^2\right)^2},0\right),\quad (A_\xi,A_\eta,A_\phi)=\left(0,0,\frac{b_m \left(\eta ^2-\xi^2\right)}{\xi^2+\eta ^2}\right)
\end{equation}
and the Hamiltonian takes the form
\begin{equation}
H=\frac{1}{2}\left(\frac{(p_\xi^A)^2+(p_\eta^A)^2}{2 \left(\eta ^2+\xi^2\right)}+\frac{(p_\phi^A)^2}{2 \eta ^2 \xi^2}\right)+\frac{ \omega }{\eta ^2+\xi^2}+\frac{2 b_m^2}{\left(\eta ^2+\xi^2\right)^2},
\end{equation}
admitting four integrals of motion
\begin{eqnarray}
\hspace{-2cm}X_1=\frac{\eta ^2 (p_\xi^A)^2- \xi^2(p_\eta^A)^2}{2 \left(\eta ^2+\xi^2\right)}+\frac{\eta^2-\xi^2}{2\xi^2\eta^2}(p_\phi^A)^2+\frac{2b_mp_\phi^A}{\eta^2+\xi^2}+\frac{ \eta ^2  \omega }{ \eta ^2+\xi^2}\nonumber\\
\hspace{-2cm}\tilde{X}_2=p_\phi^A+\frac{b_m \left(\xi^2-\eta ^2\right)}{\xi^2+\eta ^2},\nonumber\\
\hspace{-2cm}Y_3=\left(\frac{\eta^2-\xi^2}{2\eta\xi}p_\phi^A+\frac{2b_m\xi\eta }{\xi^2+\eta^2}\right)\cos (\phi )+\frac{1}{2} ( \eta p_\xi^A - \xi p_\eta^A) \sin (\phi ),\\
\hspace{-2cm}Y_4=\left(\frac{\eta^2-\xi^2}{2\eta\xi}p_\phi^A+\frac{2b_m\xi\eta }{\xi^2+\eta^2}\right)\sin (\phi )-\frac{1}{2} ( \eta p_\xi^A - \xi p_\eta^A) \cos (\phi ).\nonumber
\end{eqnarray}
This system, in the Cartesian coordinates, is described by
\begin{eqnarray}
(B_x,B_y,B_z)=\left(\frac{b_m x}{R^3},\frac{b_m y}{R^3},\frac{b_m z}{R^3}\right),\qquad R=\sqrt{x^2+y^2+z^2},\nonumber\\
(A_x,A_y,A_z)=\left(\frac{b_m yz  }{\left(x^2+y^2\right) R},\frac{-b_m xz}{\left(x^2+y^2\right) R},0\right),\nonumber\\
H=\frac{1}{2}((p_x^A)^2+(p_y^A)^2+(p_z^A)^2)+\frac{b_m^2}{2 R^2}+\frac{\omega }{2 R},\nonumber\\
X_1=L_x^Ap_y^A-L_y^Ap_x^A-\frac{b_m L_z^A}{R}- \frac{\omega z}{2R},\\
\tilde{X}_2=L_z^A+\frac{b_m z}{R},\nonumber\\
Y_3=L_x^A+\frac{b_m x}{R},\nonumber\\
Y_4=L_y^A+\frac{b_m y}{R}.\nonumber
\end{eqnarray}
This maximally superintegrable system appears in \cite{MSW15,MS17} and is characterized by the magnetic field of the magnetic monopole of the strength $b_m$ together with the Coulomb potential.

\medskip

A similar investigation for prolate and oblate spheroidal cases has so far encountered computational difficulties and we postpone it until a more efficient approach is developed.

\section{Additional integral with a leading order term $L^2$}\label{SecL2}\setcounter{equation}{0}

The results of the previous section lead us to consider also an additional quadratic integral of motion. Even in the circular parabolic case, such an investigation in full generality  seems to be too computationally challenging, thus we focus on a particular ansatz motivated by physical considerations, namely we look for a third integral of motion with the leading order term $L^2$, i.e.
\begin{equation*}
Y_3=(L_x^A)^2+(L_y^A)^2+(L_z^A)^2+\mbox{lower order terms.}
\end{equation*}
That means that our system should possess integrals which at the leading order appear like the magnitude of the angular momentum, its third component and the third component of (Laplace)--Runge--Lenz vector. In the circular parabolic coordinates, $Y_3$ takes the form
\begin{eqnarray}
Y_3=&\frac{ \xi^2(p_{\eta}^A)^2}{4}+\frac{\eta ^2 (p_{\xi}^A)^2}{4}+\frac{ \left(\eta ^2+\xi^2\right)^2(p_\phi^A)^2}{4 \eta ^2 \xi^2}\nonumber\\
&+s^\xi_3(\xi,\eta,\phi)p_\xi^A+s^\eta_3(\xi,\eta,\phi)p_\eta^A+s^\phi_3(\xi,\eta,\phi)p_\phi^A+m_3(\xi,\eta,\phi).
\end{eqnarray}

The third order compatibility conditions are automatically satisfied and the second order constrains are given by the relations
\begin{eqnarray}
p_\xi^2:\qquad & \left(\eta ^2+\xi^2\right)\partial_\xi s^\xi_3+\eta  s^\eta_3+\xi s^\xi_3=0,\nonumber\\
p_\eta^2:\qquad & \left(\eta ^2+\xi^2\right) \partial_\eta s^\eta_3+\eta s^\eta_3+\xi s^\xi_3=0,\nonumber\\
p_\phi^2:\qquad & \xi s^\eta_3+\eta s^\xi_3+\eta \xi \partial_\phi s^\phi_3=0,\\
p_\xi p_\eta:\qquad & \partial_\xi s^\eta_3+\partial_\eta s^\xi_3=0,\nonumber\\
p_\xi p_\phi:\qquad & \eta ^2 \xi^3 \left(f'\left(\eta ^2\right)+g'\left(\xi^2\right)\right)=\left(\eta ^2+\xi^2\right) \partial_\phi s^\xi_3+\eta ^2 \xi^2 \partial_\xi s^\phi_3,\nonumber\\
p_\eta p_\phi:\qquad & \eta ^3 \xi^2 \left(f'\left(\eta ^2\right)+g'\left(\xi^2\right)\right)=\left(\eta ^2+\xi^2\right) \partial_\phi s^\eta_3+\eta ^2 \xi^2 \partial_\eta s^\phi_3.\nonumber
\end{eqnarray}
A solution to this system takes the form
\begin{eqnarray}
\hspace{-2cm}f\left(\eta ^2\right)=2 b_m-\frac{1}{2} \eta ^2 \left(4 b_n+b_z \eta ^2+2 a_1\right)+a_0,\nonumber\\
\hspace{-2cm}g\left(\xi^2\right)=-\frac{b_z \xi^4}{2}+a_0+a_1 \xi^2,\nonumber\\
\hspace{-2cm}s^\xi_3(\xi,\eta ,\phi )=\frac{\eta  \sin (\phi ) \left(k_1+k_2 \left(\eta ^2+\xi^2\right)\right)-\eta  \cos (\phi ) \left(k_3+k_4 \left(\eta ^2+\xi^2\right)\right)+k_5 \xi}{\eta ^2+\xi^2},\\
\hspace{-2cm}s^\eta_3(\xi,\eta ,\phi )=\frac{\xi \sin (\phi ) \left(k_1-k_2 \left(\eta ^2+\xi^2\right)\right)+\cos (\phi ) \left(k_4 \xi \left(\eta ^2+\xi^2\right)-k_3 \xi\right)-\eta  k_5}{\eta ^2+\xi^2},\nonumber\\
\hspace{-2cm}s^\phi_3(\xi,\eta,\phi)=\frac{ \cos (\phi ) \left(k_1+k_2 \left(\eta ^2-\xi^2\right)\right)+ \sin (\phi ) \left(k_3+k_4 \left(\eta ^2-\xi^2\right)\right)}{ \eta \xi}\nonumber\\
\qquad-\left(\eta ^2+\xi^2\right) \left(\frac{b_z}{4} \left(\eta ^2+\xi^2\right)+ b_n\right).\nonumber
\end{eqnarray}
The magnetic field is given by
\begin{eqnarray}
B_\xi=\frac{4 b_m \eta  \xi^2}{\left(\eta ^2+\xi^2\right)^2}+\frac{4 b_n \eta  \xi^4}{\left(\eta ^2+\xi^2\right)^2}+b_z \eta  \xi^2,\nonumber\\
B_\eta=\frac{4 b_m \eta ^2 \xi}{\left(\eta ^2+\xi^2\right)^2}-\frac{4 b_n \eta ^4 \xi}{\left(\eta ^2+\xi^2\right)^2}-b_z \eta ^2 \xi,\label{L2mag}\\
B_\phi=0,\nonumber
\end{eqnarray}
where $b_m$, $b_n$ and $b_z$ are three arbitrary real constants parametrizing the magnetic field.

\begin{figure}[h!]
\caption{Vector field of the $b_n$ component of the magnetic field and its projection in the $xz$-plane.}
  \centering
    \includegraphics[width=0.8\textwidth]{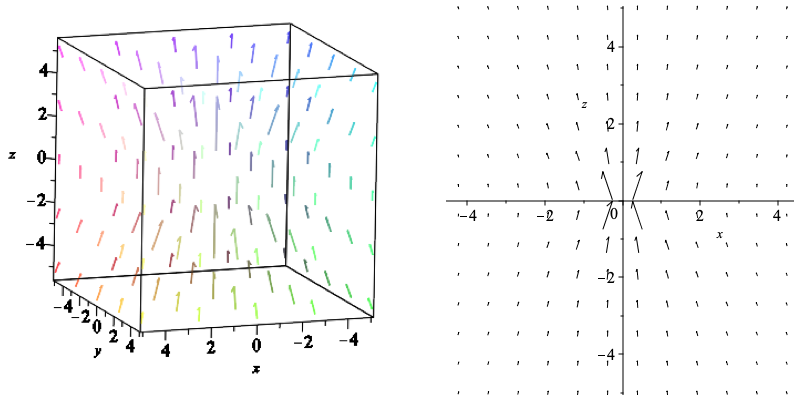}
    \label{figMag}
\end{figure}

Considering the first and zeroth order determining equations, it is convenient to first solve the dependency of $m_3$ on $\phi$ in the first order coefficient of $p_\phi$, to have only explicit functions of $\phi$. Hence, after substituting the solution in the remaining equations, we can set all the coefficients of each functionally independent function of $\phi$ to zero. Among the possible solutions, there is one interesting case which we will be exploring, i.e. the case where $k_1=k_2=k_3=k_4=k_5=0$. After gauging out some arbitrary constants, we get that 
the magnetic field is left unchanged, i.e. like in equations (\ref{L2mag}) and the vector potential can be chosen as
\begin{equation}
A_\xi=A_\eta=0,\qquad A_\phi=\frac{1}{2} b_z \eta ^2 \xi^2-\frac{2 b_m \xi^2}{\eta ^2+\xi^2}+\frac{2 b_n \eta ^2 \xi^2}{\eta ^2+\xi^2}.
\end{equation}
The Hamiltonian becomes
\begin{eqnarray}
\hspace{-1.5cm}H=&\frac{(p_r^A)^2+(p_\eta^A)^2}{2 \left(\eta ^2+\xi^2\right)}+\frac{(p_\phi^A)^2}{2 \eta ^2 \xi^2}+\frac{2 b_m^2}{\left(\eta ^2+\xi^2\right)^2}+\frac{b_m b_z \xi^2}{\eta ^2+\xi^2}-\frac{2 b_n^2 \eta ^2 \xi^2}{\left(\eta ^2+\xi^2\right)^2}-\frac{1}{8} b_z^2 \eta ^2 \xi^2\nonumber\\
\hspace{-1cm}&+b_n \left(\frac{2 b_m \left(\xi^2-\eta ^2\right)}{\left(\eta ^2+\xi^2\right)^2}-\frac{b_z \eta ^2 \xi^2}{\eta ^2+\xi^2}\right)+\frac{u_1}{\eta ^2 \xi^2}+\frac{2 u_2}{\eta ^2+\xi^2}+\frac{u_3 \left(\xi^2-\eta ^2\right)}{\eta ^2\xi^2( \xi^2+\eta ^2) },
\end{eqnarray}
where the $u_i$ are arbitrary constants appearing in the scalar potential but not in the magnetic field. This Hamiltonian admits the integrals of motion
\begin{eqnarray}
\hspace{-2cm}X_1=&\frac{\eta ^2 (p_{\xi}^A)^2-(p_{\eta}^A)^2 \xi^2}{2 \left(\eta ^2+\xi^2\right)}+\frac{1}{2}  \left(\frac{1}{\xi^2}-\frac{1}{\eta ^2}\right)(p_\phi^A)^2+b_n \left(\frac{2 b_m \eta ^2}{\eta ^2+\xi^2}+\frac{b_z \eta ^4 \xi^2}{\eta ^2+\xi^2}\right)\nonumber\\
\hspace{-2cm}&+ \left(\frac{2 b_m}{\eta ^2+\xi^2}-\frac{2 b_n \eta ^2}{\eta ^2+\xi^2}+\frac{1}{2} b_z \left(\xi^2-\eta ^2\right)\right)p_\phi^A-\frac{b_m b_z \eta ^2 \xi^2}{\eta ^2+\xi^2}+\frac{2 b_n^2 \eta ^2 \xi^2}{\eta ^2+\xi^2}\nonumber\\
\hspace{-2cm}&+\frac{1}{8} b_z^2 \eta ^2\xi^2\left(\eta ^2 - \xi^2\right)+u_1 \left(\frac{1}{\xi^2}-\frac{1}{\eta ^2}\right)+\frac{2 \eta ^2 u_2}{\eta ^2+\xi^2}-\frac{u_3 \left(\eta ^4+\xi^4\right)}{\eta ^2 \xi^2(\xi^2+\eta ^2) },\nonumber
\end{eqnarray}
\begin{equation}
\hspace{-2cm}\tilde{X}_2=p_\phi^A+\frac{2 b_m \xi^2}{\eta ^2+\xi^2}-\frac{2 b_n \eta ^2 \xi^2}{\eta ^2+\xi^2}-\frac{1}{2} b_z \eta ^2 \xi^2,
\end{equation}
\begin{eqnarray}
\hspace{-2cm}Y_3=&\frac{(p_\eta^A)^2 \xi^2}{4}+\frac{\eta ^2 (p_\xi^A)^2}{4}-\frac{1}{2} \eta  p_\eta^A p_\xi^A \xi+\frac{(p_\phi^A)^2 \left(\eta ^2+\xi^2\right)^2}{4 \eta ^2 \xi^2}\nonumber\\
\hspace{-2cm}&-p_\phi^A \left(b_n \left(\eta ^2+\xi^2\right)+\frac{1}{4} b_z \left(\eta ^2+\xi^2\right)^2\right)+\frac{1}{2} b_n b_z \eta ^2 \xi^2 \left(\eta ^2+\xi^2\right)\nonumber\\
\hspace{-2cm}&+\frac{1}{16} b_z^2 \eta ^2 \xi^2 \left(\eta ^2+\xi^2\right)^2+b_n^2 \eta ^2 \xi^2+\frac{u_1 \left(\eta ^4+\xi^4\right)}{2 \eta ^2 \xi^2}+\frac{u_3 \left(\xi^4-\eta ^4\right)}{2 \eta ^2 \xi^2}.\nonumber
\end{eqnarray}
In the Cartesian coordinates, we find
\begin{eqnarray}
B_x=&\frac{b_m x }{R^3}+\frac{b_n xz}{R^3},\nonumber\\
B_y=&\frac{b_my }{R^3}+\frac{ b_n yz}{R^3},\\
B_z=&\frac{b_m z}{R^3}+\frac{b_n \left(R^2+ z^2\right)}{R^3}+b_z.\nonumber
\end{eqnarray}
and
\begin{eqnarray}
A_x=&\frac{b_m y z}{\left(x^2+y^2\right) R}-\frac{b_n y}{ R}-\frac{b_z y}{2},\nonumber\\
A_y=&-\frac{b_m x z}{\left(x^2+y^2\right) R}+\frac{b_n x}{ R}+\frac{b_{z} x}{2},\\
A_z=&0.\nonumber
\end{eqnarray}
Thus the magnetic field generated by the constant $b_z$ represents a constant magnetic field in the $z$-direction and the magnetic field generated by the constant $b_m$ takes the form of the magnetic field of the magnetic monopole. The magnetic field coming from the constant $b_n$ is illustrated in figures \ref{figMag}. This magnetic field is bounded by the relation $\vert\vec{B}\vert\leq\frac{2\vert b_n\vert}{R}$, when $b_z$ and $b_m$ are set to zero.

The integrals of motion become
\begin{eqnarray}
\hspace{-2cm}H=&\frac{(p_x^A)^2}{2}+\frac{(p_y^A)^2}{2}+\frac{(p_z^A)^2}{2}+\frac{u_1}{x^2+y^2}+\frac{u_2}{R}+\frac{u_3 z}{\left(x^2+y^2\right) R}\nonumber\\
\hspace{-2cm}&+\frac{b_m^2}{2 R^2}+ \frac{b_zb_m z}{2 R}-\frac{b_zb_n \left(x^2+y^2\right)}{2 R}+\frac{b_m b_n z}{R^2}-\frac{ b_n^2(x^2+y^2)}{2R^2}-\frac{1}{8} b_z^2 \left(x^2+y^2\right),\nonumber
\end{eqnarray}
\begin{eqnarray}
\hspace{-2cm}X_1=&p_y^AL_x^A-p_x^AL_y^A+L_z^A \left(\frac{b_m }{ R}+\frac{b_n  z}{ R}+ b_z  z\right)-\frac{b_m b_z \left(x^2+y^2\right)}{2 R}\nonumber\\
\hspace{-2cm}&-\frac{b_n b_z z \left(x^2+y^2\right)}{2 R}-\frac{b_z^2 z}{4} \left(x^2+y^2\right)-\frac{2u_1 z}{x^2+y^2}-\frac{u_2 z}{ R}-\frac{u_3 \left(R^2+ z^2\right)}{ \left(x^2+y^2\right) R},\nonumber
\end{eqnarray}
\begin{eqnarray}
\hspace{-2cm}\tilde{X}_2=&L_z^A+\frac{b_m z}{R}-\frac{b_n \left(x^2+y^2\right)}{R}-\frac{b_z}{2}  \left(x^2+y^2\right),
\end{eqnarray}
\begin{eqnarray}
\hspace{-2cm}Y_3=&(L^A)^2-L_z^A \left(2 b_n  R+b_z  R^2\right)+b_n^2 \left(x^2+y^2\right)\nonumber\\
\hspace{-2cm}&+b_n b_z \left(x^2+y^2\right) R+\frac{1}{4} b_z^2 \left(x^2+y^2\right) R^2+\frac{2 u_1 z^2}{x^2+y^2}+\frac{2 u_3 z R}{x^2+y^2}.\nonumber
\end{eqnarray}
This (minimally) superintegrable system wasn't encountered in the context of superintegrability yet and admits a new form of magnetic field coming from the coefficient $b_n$. 

This superintegrable system can be alternatively obtained from the spherical case by looking for an additional (Laplace)--Runge--Lenz type integral of motion. The same constraints come up since both the $L^2$-type and (Laplace)--Runge--Lenz-type integrals of motion Poisson-bracket commute with the $L_z^2$-type integral of motion, i.e.
\begin{equation}
\lbrace Y_3,X_2\rbrace=\lbrace X_1,X_2\rbrace=0.
\end{equation}
Hence, the same constrains are imposed a posteriori.

The Poisson bracket of $X_1$ and $Y_3$ does not vanish, but its square can be expressed in terms of $X_1$, $\tilde{X}_2$, $Y_3$ and $H$, as follows
\begin{eqnarray}
\hspace{-2cm}\left(\lbrace X_1,Y_3\rbrace\right)^2=&-8  \tilde{X}_2^2 Y_3 H + 8 (Y_3)^2 H - 4 X_1^2 Y_3 + 4 b_z  \tilde{X}_2^3 Y_3 -  4 b_z  \tilde{X}_2 (Y_3)^2\nonumber\\
&+ 4 b_n^2  \tilde{X}_2^2 Y_3 + 8 (b_m^2+2u_1) H Y_3 + 8 b_n b_m Y_3 X_1\nonumber\\
& - 8 b_n b_m  \tilde{X}_2^2 X_1 -  4 b_n^2  \tilde{X}_2^4 - 4 (b_m^2+2u_1) X_1^2 - 16 u_1  \tilde{X}_2^2 H \nonumber\\
&  - 16 b_m u_3  \tilde{X}_2 H - 8 (b_m u_2 - b_n u_3)  \tilde{X}_2 X_1\nonumber\\
& +  8 (b_z u_1 - b_n u_2)  \tilde{X}_2^3 +  4 (2 b_n u_2 - b_m^2 b_z - 2 b_z u_1)  \tilde{X}_2 Y_3+ 8 u_3^2 H \nonumber\\
& +  8 (b_m^3 b_n + 2 b_m b_n u_1 + u_2 u_3) X_1 +  4 (2 b_m^2 b_n^2 - u_2^2 + 2 b_m b_z u_3)  \tilde{X}_2^2\nonumber\\
& -  4 (b_m^2 b_n^2 - u_2^2) Y_3 +  4 (2 b_m^2 b_n u_2 - 2 b_m b_n^2 u_3 - b_z u_3^2)  \tilde{X}_2\nonumber\\
& -  4 b_m b_n (b_m^3 b_n + 2 b_m b_n u_1 + 2 u_2 u_3)
\end{eqnarray}
Hence, the algebra of the known integrals of motion closes polynomially and doesn't contain any additional independent integral.

\begin{figure}[h!]
\caption{Trajectory for the values $ b_z=-2/7$, $b_m=-1/2$, $b_n=-5/2$, $u_1=1/6$, $u_2=-3/2$, $u_3=0$ with the initial conditions $x(0)=1$, $y(0)=0$, $z(0)=0$, $p_x(0)=0$, $p_y(0)=1$, $p_z(0)=1/2$.}
  \centering
    \includegraphics[width=0.7\textwidth]{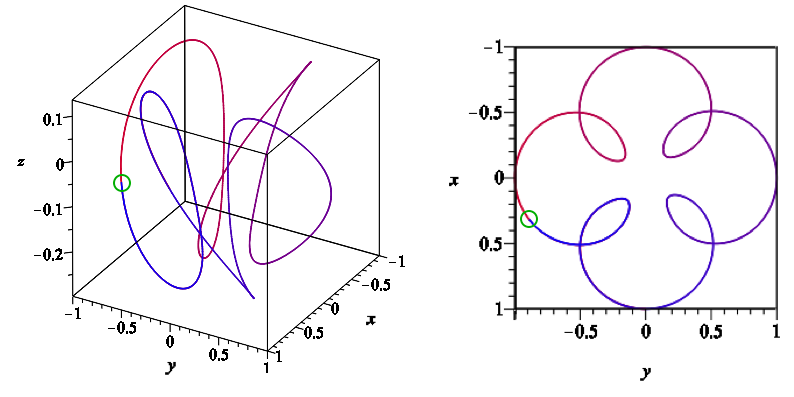}
    \label{Ex1}
\end{figure}

\begin{figure}[h!]
\caption{Trajectory for the values $ b_z=-2/7$, $b_m=-5/2$, $b_n=-5/2$, $u_1=1/6$, $u_2=-3/2$, $u_3=0$ with the initial conditions $ x(0)=1$, $y(0)=0$, $z(0)=0$, $p_x(0)=0$, $p_y(0)=1$, $p_z(0)=1/2$.}
  \centering
    \includegraphics[width=0.7\textwidth]{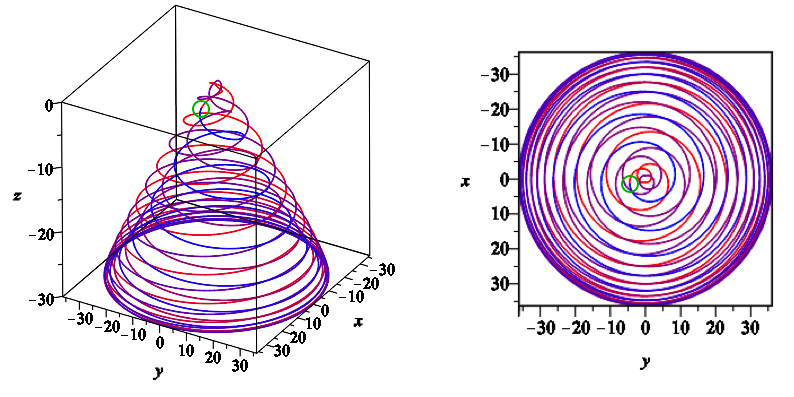}
    \label{Ex2}
\end{figure}

\begin{figure}[h!]
\caption{Trajectory for the values $b_z=0$, $b_m=0$, $b_n=-2$, $u_1=1/2$, $u_2=-1$, $u_3=-1/4$ with the initial conditions $ x(0)=1$, $y(0)=0$, $z(0)=0$, $p_x(0)=0$, $p_y(0)=1$, $p_z(0)=1/2$.}
  \centering
    \includegraphics[width=0.7\textwidth]{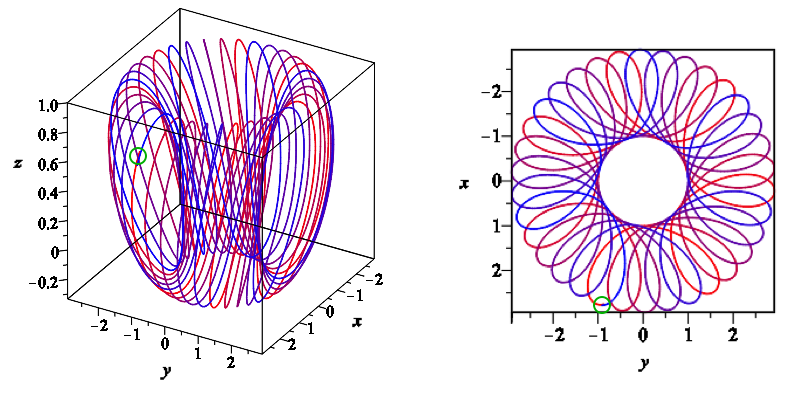}
    \label{Ex3}
\end{figure}

By looking for numerical approximations of the trajectories of such a system, we notice that all bounded trajectories seem to be closed for randomly chosen values of the magnetic field and potential, as one can see in figures \ref{Ex1} to \ref{Ex3}. This property suggests that the system may be maximally superintegrable, i.e. this Hamiltonian system may possess an additional integral of motion that we didn't find yet. (Time evolution of the trajectories in the figures \ref{Ex1} to \ref{Ex3} is indicated by the change of colour, i.e. red implies $t=0$ and it gets blue as time is moving forward. The small green circle indicates the position of the (apparent) closing of the trajectory.)

For a hypothetical fifth independent quadratic integral of the form (\ref{genx}), we considered the compatibility conditions for functions $s_j$, which are second order partial differential equations involving the magnetic field and the constants $\alpha_{jk}$. After the elimination of the known integrals, these imply vanishing of all remaining $\alpha_{jk}$ (or of the constant $b_n$), i.e. no independent second order integral exists for $b_n$ non-vanishing. However, this observation does not exclude existence of an integral of higher order, or even a non-polynomial one. We presently don't know how to find it if it exists.

\pagebreak

\section{Conclusions}\label{SecConc}\setcounter{equation}{0}
We have investigated all 3D systems with magnetic field whose structure of quadratic integrals of motion corresponds to three non-subgroup type coordinate systems distinguished by their rotational invariance around one coordinate axis. We have found that the results are significantly different from the subgroup type, namely Cartesian, spherical and cylindrical, cases.

Namely,
\begin{itemize}
\item  for all systems with subgroup type integrals, there exist integrable systems which possess two truly second order integrals of motion, see \cite{Zhalij15} for the Cartesian type, \cite{MSW18} for the spherical type and a simple extension of 2D systems of~\cite{MW00} into the $z$-direction by adding $\frac{1}{2} p_z^2+V(z)$ to the potential for the cylindrical type (or work in progress on cylindrical type integrals), respectively. However, in all non-subgroup type integrable systems considered in this paper, the assumption of existence of two commuting integrals of the given form implies the existence of a first order integral $\tilde{X}_2$ which in the chosen gauge reduces to the angular momentum component
\begin{equation}
\tilde{X}_2=L_z
\end{equation}
and implies rotational invariance of the constructed systems. 
\item We notice that the explicit form of the integrable systems constructed under the assumption of the existence of the commuting quadratic integrals of the given form exactly matches the systems constructed by Benenti, Chanu and Rastelli in~\cite{BCR01}. Thus, the necessary existence of the first order integral $L_z$ implies separability of the Hamilton--Jacobi equation and we recover for these non-subgroup type integrals the equivalence of quadratic integrability and separability, familiar from the consideration of the systems without magnetic field. 
\item We have seen that the structure of the magnetic field and the first order coefficients $s_a^j$ is fully determined by the highest order determining equations and is not in any way affected by the lower order conditions - these in turn determine the possible structure of the scalar potential and the lowest order terms in the integrals.

This is not the case for any of the subgroup type systems where lower order conditions through their compatibility always further constrain the magnetic field.
\end{itemize}

Concerning superintegrability, out of numerous possible directions we have so far systematically analyzed only two, both for the circular parabolic case: the first order superintegrability, for which no new system exists, and the second order superintegrability with $L^2$-type integral. In this case, an interesting new system was found and some of its properties discussed, like apparent presence of closed bounded trajectories. It remains an open problem to establish or exclude its hypothetical higher order maximal superintegrability and its potential physical interest, e.g. in plasma physics where its superintegrability and thus expected certain resilience with respect to perturbations may be of importance.
\medskip

Our paper invokes many open questions and directions for further research.

First of all, it would be of importance to establish or disprove the equivalence of quadratic integrability and separability of Hamilton--Jacobi equations also for other non-subgroup type pairs of commuting integrals, i.e. the ones which do not possess $L_z^2$-type integral. However, we expect serious computational difficulties in this direction, in particular in ellipsoidal and paraboloidal cases, and we are not yet sure that for these cases a similar investigation is feasible. 

Secondly, some more efficient techniques need to be developed to deal with quadratic superintegrability in full generality, not just using a particular ansatz like in section \ref{SecL2}. Without some new ideas a full classification of all possible second order integrals for the given class of integrable systems with magnetic fields seems to be presently out of reach.

Thirdly, superintegrability in the oblate and prolate spheroidal cases should be investigated, both the first and second order one. Even at the first order level this appears to be computationally significantly more challenging compared to the circular parabolic case.

Fourthly, in most potential physical applications relativistic velocities may be reached, thus a relativistic version of the system of section \ref{SecL2} should be also investigated, like it was done in \cite{HI17} for some other superintegrable systems.

Last but not least, our analysis in this paper was purely classical, the quantum version of the problem was not considered at all. The quantum corrections are currently under investigation and we postpone them to later work.

\section*{Acknowledgements}
SB was supported by a postdoctoral fellowship provided by the Fonds de Recherche du Qu\'ebec : Nature et Technologie (FRQNT). The research of L\v{S} was supported by the Czech Science Foundation (GACR), project 17-11805S. The authors thank Antonella Marchesiello and Pavel Winternitz for discussions on the subject of this paper.

%\bibliography{3DMagnbib}
%\bibliographystyle{unsrt}
%\bibliographystyle{plain}

\end{document}